# Indirect search for light charged Higgs bosons through the dominant semileptonic decays of top quark $t \to b(\to B/D + X) + H^+(\to \tau^+ \nu_\tau)$

S. Abbaspour [a], S. Mohammad Moosavi Nejad [a,b,∗], M. Balali [a]

[a] *Faculty of Physics, Yazd University, P.O. Box 89195-741, Yazd, Iran*
[b] *School of Particles and Accelerators, Institute for Research in Fundamental Sciences (IPM), P.O.Box 19395-5531, Tehran, Iran*

**Abstract**

In this work we introduce a new channel to indirect search for the light charged Higgs bosons, which are predicted in several extensions of the standard model (SM) such as the two-Higgs-doublet models (2HDMs). We calculate the $\mathcal{O}(\alpha_s)$ QCD radiative corrections to the energy distribution of bottom- and charmed-flavored hadrons ($B/D$) produced in the dominant decays of the polarized top quark in the 2HDM, i.e. $t(\uparrow) \longrightarrow b(\to B/D + \text{jet}) + H^+(\to \tau^+ \nu_\tau)$. Generally, the energy distribution of hadrons is governed by the unpolarized rate and the polar and the azimuthal correlation functions which are related to the density matrix elements of the decay $t(\uparrow) \to bH^+$. In our proposed channel, any deviation of the $B/D$-meson energy spectrum from its corresponding SM predictions can be considered as a signal for the existence of charged Higgs at the LHC. We also calculate, for the first time, the azimuthal correlation rate $\Gamma_\phi$ at next-to-leading order which vanishes at the Born level.

---

\* Corresponding author.
   *E-mail address:* mmoosavi@yazd.ac.ir (S.M. Moosavi Nejad).

## 1. Introduction

Charged Higgs bosons emerge in the scalar sector of several standard model (SM) extensions, and are the object of various beyond SM searches at the CERN Large Hadron Collider (LHC). Since the SM does not include any elementary charged scalar particle, then the experimental observation of a charged Higgs boson would necessarily be a signal for a nontrivially extended scalar sector and a definitive evidence of new physics beyond SM. In recent years, searches for charged Higgs bosons have been done by the ATLAS and the CMS collaborations at the LHC in proton-proton collision and numerous attempts are still in progress.

Among many beyond SM scenarios which motivate the existence of charged Higgs, a generic two-Higgs-doublet model (2HDM) [1–3] provides a greater insight of the SUSY Higgs sector without including the plethora of new particles which SUSY predicts. Within this class of models, two isospin doublets are introduced to break the symmetry of $SU(2) \times U(1)$. This symmetry breaking leads to the existence of five physical Higgs bosons; three physical neutral Higgs bosons (h, H, A) and a pair of charged-Higgs bosons ($H^{\pm}$) [2].

The dominant production and decay modes for a charged Higgs depend on the value of its mass with respect to the top-quark mass, and can be classified into three categories [4]. Light charged Higgs scenarios are defined by Higgs-boson masses smaller than the top quark mass. In the 2HDM, the main production mode for light charged Higgses is through the top quark decay $t \to bH^+$. Therefore at the LHC, as an formidable machine producing around 90 million top pairs per year of running at design c.m. energy of 14 TeV, the light Higgs bosons can be searched in the subsequent decay products of the top pairs $t\bar{t} \to H^{\pm}H^{\mp}b\bar{b}$ and $t\bar{t} \to H^{\pm}W^{\mp}b\bar{b}$ when $H^{\pm}$ decays into $\tau$ lepton and neutrino.

On the other hand, in the decay mode $t \to bH^+$ both b-quarks hadronize into the b-jet $X_b$ before they decay and the Higgs bosons decay into the leptons and their related neutrinos, i.e. $H^+ \to l^+\nu_l (l = e, \mu, \tau)$. Therefore, at the LHC the decay process $t \to b(\to X_b + \text{jet}) + H^+(\to l^+\nu_l)$ is of prime importance and it is an urgent task to predict its partial decay width as reliably as possible. Of particular interest are the determination of energy distribution of hadrons inclusively produced in the top quark rest frame. The study of these hadronic energy distributions in the polarized and unpolarized top decays could be proposed as a new approach to indirect search for the charged Higgs bosons at the LHC. Practically, any deviation of the energy spectrum of the produced hadrons in top quark decays from the SM predictions can be considered as a signal for the existence of charged Higgs. In [5], the energy distribution of bottom-flavored hadrons (B) inclusively produced in the SM decay chain of an unpolarized top quark, i.e. $t \to bW^+ \to Bl^+\nu_l + X$, is studied and in [6] the energy distribution of B-hadrons is investigated in the unpolarized top decays in the 2HDM scenarios, i.e. $t \to bH^+ \to BH^+ + X$, where $X$ refers to the unobserved final state particles.

Since, the life time of the top quark ($\approx 4.6 \times 10^{-25}$ s) is much shorter than the typical time needed for the QCD interactions to randomize its spin, therefore, its full polarization content is preserved and passes on to its decay products, so that the polarization of the top quark will reveal itself in the angular decay distribution. Thus, the top quark polarization can be studied through the angular correlations between the direction of the top quark spin and the momenta of its decay products [7].

In this work, in the 2HDM framework we study the $\mathcal{O}(\alpha_s)$ angular distribution of energy spectrum of B/D-mesons considering the polar and the azimuthal angular correlations in the rest frame decay of a polarized top quark, i.e. $t(\uparrow) \to B/D + H^+ + X$ followed by $H^+ \to l^+\nu_l$. This angular correlation is analyzed in a specific helicity coordinate system where the event plane, including the top and its decay products, is defined in the $(\hat{x}, \hat{z})$ plane and the b-quark three-momentum points into the direction of the positive $\hat{z}$-axis. In this frame, the top quark polarization vector is evaluated with respect to the direction of the b-quark three-momentum. To define the event plane, one needs to measure the momentum directions of the b-quark and the $H^+$-boson and the polarization direction of the top quark. The evaluation of the b-quark momentum direction requires the use of a jet finding algorithm, while the top spin direction must be obtained from theoretical input, e.g. a polarized linear $e^+e^-$ collider may be considered as a copious source of close to zero and close to 100% polarized tops.

The azimuthal correlation between the event plane and the intersecting one to this plane (including the top quark polarization vector) in the semileptonic rest frame decay of a polarized top quark (see Fig. 3) belongs to a class of polarization observables involving the top quark in which the leading-order contribution receives a zero result. As we will show, the nonzero contributions can arise from higher order radiative corrections. Due to the zero result for the lowest order contribution, the azimuthal decay rate up to NLO will be small. Then, it seems that the measurement of the azimuthal correlation would be difficult, but since highly polarized top quarks with more accuracy will become available at higher luminosity hadron colliders through single top production processes [8], it might then be feasible to experimentally measure this azimuthal correlation through the energy distribution of hadrons from polarized top decays. As was explained, the comparison of these polar and azimuthal distributions of meson energy spectrum produced in polarized top decays with the SM ones might be considered as a new channel to indirect search for the charged Higgs bosons at the LHC.

Concerning the importance and other applications of the analytical results presented in this work, note that, however current ATLAS and CMS measurements exclude a light charged Higgs for most of the parameter regions in the context of the minimal supersymmetric standard model (MSSM) scenarios, these bounds are significantly weakened in the Type II 2HDM (MSSM) once the exotic decay channel into a lighter neutral Higgs, $H^\pm \to AW^\pm/HW^\pm$, is open. In [9], the possibility of a light charged Higgs produced in top decay via single top or top pair production is examined with a subsequent decay as $H^\pm \to AW^\pm/HW^\pm$. It is shown that, this decay can reach a sizable branching fraction at low $\tan\beta$ once it is kinematically permitted. Their results show that the exotic decay channel $H^+ \to AW^+/HW^+$ is therefore complementary to the conventional $H^+ \to \tau^+\nu_\tau$ channel, considered in the current MSSM scenarios. Therefore, our analytical results presented in this work will be also able to be applied for the decay process $t(\uparrow) \to b(\to B/D + jet) + H^+(\to AW^+/HW^+)$ considering a convenient branching fraction for the decay $H^+ \to AW^+/HW^+$. Note that, due to experimental challenges at low energies, such a light neutral Higgs has not been fully excluded yet.

## 2. Unpolarized top quark decay in the narrow width approximation

From the unitarity of the Cabibbo–Kobayashi–Maskawa (CKM) quark mixing matrix [10], one has the relation

$$\overbrace{|V_{ub}|^2}^{(\approx 0.004)^2} + \overbrace{|V_{cb}|^2}^{(\approx 0.04)^2} + |V_{tb}|^2 = 1, \tag{1}$$

so, it results $|V_{tb}| \approx 1$ to a very high accuracy. Therefore, top quark decays within the SM are completely dominated by the mode $t \to bW^+$ followed by $W^+ \to l^+\nu_l$ ($l = e, \mu, \tau$). In theories beyond SM with an extended Higgs sector, assuming $m_t > m_b + m_{H^+}$, one may also have the decay mode $t \to bH^+$ followed by $H^+ \to l^+\nu_l$. Although, as an example, the $\tau^+$ leptons arising from the decays $W^+ \to \tau^+ \nu_\tau$ and $H^+ \to \tau^+ \nu_\tau$ are predominantly left- and right-polarized, respectively. The polarization of the $\tau^+$ influences the energy distributions of the decay products in the subsequent decays of the $\tau^+ \to \pi^+ \bar{\nu}_\tau, \rho^+ \bar{\nu}_\tau, l^+ \nu_l \bar{\nu}_\tau$. These distributions are not discussed in our work. For more detail, see Ref. [11].

Here, we first review some technical detail about the decay rate of unpolarized top quarks in the process

$$t \to b + H^+ \to b + (l^+ \nu_l), \tag{2}$$

by working in the general two Higgs doublet model, where $H_1$ and $H_2$ are the doublets whose vacuum expectation values (VEVs) give masses to the down and up type quarks. Considering the VEVs of the fields $H_1(\mathbf{v}_1)$ and $H_2(\mathbf{v}_2)$, one has the constraint $\mathbf{v}_1^2 + \mathbf{v}_2^2 = (\sqrt{2}G_F)^{-1}$ where $G_F$ is the Fermi's constant which is related to the weak coupling factor as $G_F = g_W^2 / (4\sqrt{2} m_W^2)$. The ratio of VEVs is a free parameter and is defined as $\tan \beta = \mathbf{v}_2 / \mathbf{v}_1$. Even, a linear combination of the charged components of $H_1$ and $H_2$ gives two massive charged Higgs boson states $H^\pm$, i.e. $H^\pm = H_2^\pm \cos \beta - H_1^\pm \sin \beta$.

In a general 2HDM, in order to avoid tree-level flavor-changing neutral currents that can be induced by Higgs-boson exchange, the generic Higgs coupling to all quarks should be restricted. Fortunately, there are several classes of 2HDMs which naturally avoid this difficulty by restricting the Higgs coupling. Imposing flavor conservation, there are four possible ways (hereinafter called models I–IV) for the two Higgs doublets to couple to the SM fermions. Each of these four ways gives rise to rather different phenomenology predictions. In these models, the generic charged Higgs coupling to fermions (assuming massless neutrinos) can be expressed as a superposition of right- and left-chiral coupling factors [12], so that the relevant part of the interaction Lagrangian of the process (2) is given by

$$\begin{aligned} L_I = \frac{g_W}{2\sqrt{2} m_W} H^+ \Big\{ & V_{tb} \big[ \bar{u}_t(p_t) \{ A(1 + \gamma_5) \\ & + B(1 - \gamma_5) \} u_b(p_b) \big] \\ & + C \big[ \bar{u}_{\nu_l}(p_\nu)(1 - \gamma_5) u_l(p_l) \big] \Big\}, \end{aligned} \tag{3}$$

where A, B and C are model-dependent parameters.

In the first model (model I), one doublet $H_2$ gives masses to all quarks and leptons and the other doublet $H_1$ essentially decouples from fermions. In this model, one has

$$A_I = m_t \cot \beta, \quad B_I = -m_b \cot \beta, \quad C_I = -m_\tau \cot \beta. \tag{4}$$

In a second model (model II), the doublet $H_2$ gives mass to the right-chiral up-type quarks (and possibly neutrinos) and another doublet $H_1$ gives mass to the right-chiral down-type quarks and charged leptons. In this model, the interaction Lagrangian (3) includes

$$A_{II} = m_t \cot \beta, \quad B_{II} = m_b \tan \beta, \quad C_{II} = m_\tau \tan \beta. \tag{5}$$

There are also two other possibilities (models III and IV) in which the down-type quarks and charged leptons receive mass from different doublets; in model III both up- and down-type quarks couple to the second doublet ($H_2$) and all leptons to the first one, so one has

$$A_{III} = m_t \cot \beta, \quad B_{III} = m_b \tan \beta, \quad C_{III} = -m_\tau \cot \beta \tag{6}$$

and in model IV, the roles of the two doublets are reversed with respect to the model II, so that

$$A_{IV} = m_t \cot\beta, \quad B_{IV} = -m_b \cot\beta, \quad C_{IV} = m_\tau \tan\beta. \tag{7}$$

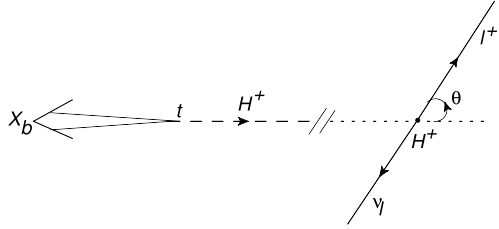

Fig. 1. Lowest order contribution to $t \to b + H^+ (\to l^+ \nu_l)$. The polar angle $\theta$ is defined in the $H^+$-rest frame.

These models are also known as type I–IV 2HDM scenarios. The type-II scenario is, in fact, the Higgs sector of the MSSM up to SUSY corrections [13,14]. In other words, in the MSSM one has a type-II 2HDM sector in addition to the supersymmetric particles, in particular the charginos, stops and gluinos. In this work to present our phenomenological predictions for the energy distribution of produced hadrons we restrict ourselves to the MSSM scenario, although we will discuss how our results can be generalized to other types.

We start by discussing the Born term contribution to the decay rate of the process $t \to b l^+ \nu_l$ ($l^+ = e^+, \mu^+, \tau^+$) (2), where the top quark is presently considered as unpolarized. We consider the decay process

$$t(p_t) \to b(p_b) + H^+(p_{H^+}) \to b(p_b) + l^+(p_l) + \nu_l(p_\nu), \tag{8}$$

where the four-momentum assignments are indicated in parentheses, see also Fig. 1. Here, to be specific, we concentrate on the case with $l^+ \nu_l = \tau^+ \nu_\tau$. Using the couplings from the Lagrangian (3) one can write the matrix element of the process (8) in the following form

$$\begin{aligned}M_0(t \to b\tau^+\nu_\tau) &= i\frac{g_W^2 V_{tb}}{8m_W^2} \frac{1}{p_{H^+}^2 - m_{H^+}^2 + im_{H^+}\Gamma_{H^+}} \\ &\times C[\bar{u}_\nu(p_\nu)(1+\gamma_5)v_\tau(p_\tau)] \\ &\times [\bar{u}_t(p_t)\{A(1-\gamma_5) + B(1+\gamma_5)\}u_b(p_b)].\end{aligned} \tag{9}$$

Note that, since the main contribution to the decay mode (8) comes from the kinematic region where $H^+$ boson is near its mass-shell, then one has to take into account its finite decay width $\Gamma_{H^+}$. For this reason, in (9) we used the Breit–Wigner prescription of the Higgs propagator. In (9), the model dependent parameters A, B, C are defined in (4)–(7). From now on and for simplicity we introduce

$$\begin{aligned}a &= \frac{g_W}{2\sqrt{2}m_W}|V_{tb}|(A+B), \\ b &= \frac{g_W}{2\sqrt{2}m_W}|V_{tb}|(A-B), \\ c &= \frac{g_W}{2\sqrt{2}m_W}C.\end{aligned} \tag{10}$$

By this, the coupling of the charged Higgs to the bottom and top quarks is expressed as a superposition of scalar and pseudoscalar coupling factors.

On squaring the Born matrix element (9) and taking the spin sums, one is led to the Born contribution as

$$|M_0|^2 = \frac{1}{(p_{H^+}^2 - m_{H^+}^2)^2 + (m_{H^+}\Gamma_{H^+})^2} \\ \times |M^{Born}(t \to bH^+)|^2 \\ \times |M^{Born}(H^+ \to \tau^+ \nu_\tau)|^2, \quad (11)$$

where $|M^{Born}(H^+ \to \tau^+ \nu_\tau)|^2 = 8c^2(p_\nu \cdot p_\tau)$ and $|M^{Born}(t \to bH^+)|^2 = 4[(a^2-b^2)m_b m_t + (a^2+b^2)p_b \cdot p_t]$. For the 3-body decay rate $t \to b\tau^+\nu_\tau$, one has

$$d\Gamma^{Born} = \frac{1}{2m_t} \prod_{i=b,\tau^+,\nu_\tau} \frac{1}{(2\pi)^3} \frac{d^3 p_i}{2E_i} \overline{|M_0|^2} (2\pi)^4 \delta^4(p_t - \sum_{i=b,\tau^+,\nu_\tau} p_i), \quad (12)$$

where $\overline{|M_0|^2} = |M_0|^2/(1+2s_t)$ for which $s_t$ stands for the top quark spin. By working in the narrow-width approximation (NWA), the Breit–Wigner Resonance is replaced by a delta-function as [15]

$$\frac{1}{(p_{H^+}^2 - m_{H^+}^2)^2 + (m_{H^+}\Gamma_{H^+})^2} \\ \approx \frac{\pi}{m_{H^+}\Gamma_{H^+}} \delta(p_{H^+}^2 - m_{H^+}^2). \quad (13)$$

Using the NWA for the $H^+$-boson, the three body decay $t \to b\tau^+\nu_\tau$ is factorized as

$$\Gamma(t \to b\tau^+\nu_\tau) = \Gamma(t \to bH^+) \frac{\Gamma(H^+ \to \tau^+\nu_\tau)}{\Gamma_{H^+}} \\ = \Gamma(t \to bH^+) B_\tau^H(H^+ \to \tau^+\nu_\tau), \quad (14)$$

which is a result expected from physical intuition.

One considerable point about the interference terms which are ignored in our work: in this work, using the NWA we study the decay process $t(\uparrow) \to bH^+$ in the 2HDM scenario so in [16] we used the same approximation for the SM decay one, i.e. $t(\uparrow) \to bW^+$. An important condition limiting the applicability of this approximation, however, is the requirement that there should be no interference of the contribution of the intermediate particle for which the NWA is applied with any other close-by resonance. Indeed, if the mass gap between two intermediate particles is smaller than one of their total widths, the interference term between the contributions from the two nearly mass-degenerate particles may become large. In other words, interference effects can be large if there are several resonant diagrams whose intermediate particles (with masses $M_1$ and $M_2$ for two resonances) are close in mass compared to their total decay widths: $|M_1 - M_2| \leq (\Gamma_1, \Gamma_2)$ [15]. In these cases, a single-resonance approach or the incoherent sum of two resonance contributions does not necessarily hold. Then, if the mass difference is smaller than their total widths, the two resonances overlap. This can lead to a potentially large interference term, which is neglected in the standard NWA, but can be taken into account in the full calculation or in a generalized NWA [15]. In our calculations, the required condition to apply the NWA holds, i.e. $|m_{H^+} - m_{W^+}| \gg (\Gamma_{W^+}, \Gamma_{H^+})$ if one sets $\Gamma_W = 2.212$ GeV (with $m_W = 80$ GeV) and $\Gamma_{H^+} \leq 0.08$ (for $m_{H^+} = 95$ and 160 GeV which are applied in our work). Therefore, the overlap of two resonances can be ignored with high accuracy and two decay modes $t \to bH^+ \to b\tau^+\nu_\tau$ and $t \to bW^+ \to b\tau^+\nu_\tau$ can be studied separately, so the total decay width of top quarks can be simply obtained by the summation of both decay rates. More details about the interference effects in BSM processes with a generalized NWA can be found in [15].

## 2.1. Born level results for unpolarized top decays

In this section we calculate the Born level rate of the process (8) using the NWA where we put $p_{H^+}^2 = m_{H^+}^2$ from the beginning. According to Eq. (14), the Born width is expressed as

$$\Gamma_0(t \to b\tau^+ \nu_\tau) = \Gamma_0(t \to bH^+) \frac{\Gamma_0(H^+ \to \tau^+ \nu_\tau)}{\Gamma_{H^+}^{\text{total}}}. \tag{15}$$

To proceed, let us introduce the following notation for the kinematic variables

$$R = \frac{m_b^2}{m_t^2}, \quad y = \frac{m_{H^+}^2}{m_t^2}, \quad S = \frac{1 + R - y}{2},$$

$$Q = \sqrt{S^2 - R}, \quad \eta = \frac{\sqrt{R}}{S}. \tag{16}$$

The Born level amplitude for the process $t \to bH^+$ is parametrized as $M_0 = \bar{u}_b(a + b\gamma_5)u_t$, so for the amplitude squared one has

$$|M_0|^2 = \sum_{s_t, s_b} M_0^\dagger M_0 = 2(p_b \cdot p_t)(a^2 + b^2) + 2(a^2 - b^2) m_b m_t, \tag{17}$$

therefore, the tree-level decay width reads

$$\tilde{\Gamma}_0(t \to bH^+) = \frac{m_t}{16\pi} \Big\{ (a^2 + b^2)(1 + R - y) + 2(a^2 - b^2)\sqrt{R} \Big\} \lambda^{\frac{1}{2}}(1, R, y), \tag{18}$$

where $\lambda(x, y, z) = (x - y - z)^2 - 4yz$ is the triangle function and the coefficients "a" and "b" are defined in (10). The NLO QCD radiative corrections to the unpolarized rate (18) are given in our previous work [6].

Next, for the squared amplitude of the leptonic decay of charged Higgs one has:

$$\overline{|M_{H^+ \to \tau^+ \nu_\tau}^{\text{Born}}|^2} = 8c^2 p_\nu \cdot p_\tau.$$

This rate must be evaluated in the top quark rest frame. In the $H^+$-rest frame ($H_{r.f.}$) shown in Fig. 1, one has $p_\tau^\mu(H_{r.f.}) = (m_{H^+}/2)[1; \sin\theta, 0, \cos\theta]$, where the angle $\theta$ is defined in the $H^+$-boson rest frame. To obtain the lepton four-momentum in the top rest frame, the relevant Lorentz boost matrix reads

$$L(\text{boost}) = \frac{1}{m_{H^+}} \begin{pmatrix} E_{H^+} & 0 & 0 & |\vec{p}_{H^+}| \\ 0 & m_{H^+} & 0 & 0 \\ 0 & 0 & m_{H^+} & 0 \\ |\vec{p}_{H^+}| & 0 & 0 & E_{H^+} \end{pmatrix} \tag{19}$$

where $E_{H^+} = m_t(1 - S)$ and $|\vec{p}_{H^+}| = m_t Q$. Therefore, the lepton four-momentum in the top rest frame reads

$$p_\tau^\mu = L(\text{boost}) p_\tau^\mu(H_{r.f.}) \tag{20}$$
$$= \frac{m_t}{2}(1 - S + Q\cos\theta; \sqrt{y}\sin\theta, 0, Q + (1 - S)\cos\theta).$$

Thus $p_\nu \cdot p_\tau = m_t^2[(1-S)^2 - Q^2]/2 = m_{H^+}^2/2$. Considering the notation defined in (16), the tree-level decay width for the leptonic sector of (8) is given by

$$\Gamma_0(H^+ \to \tau^+ \nu_\tau) = \frac{m_{H^+}}{4\pi} c^2, \quad (21)$$

which is a dependent-model rate.

Since, current search strategies assume that the charged Higgs decays either leptonically ($H^+ \to \tau^+ \nu_\tau$) or hadronically ($H^+ \to c\bar{s}$), then following Ref. [17] we adopt the relevant branching fraction $B_\tau^H(H^+ \to \tau^+ \nu_\tau)$ (14) as

$$B_\tau^H = \frac{\Gamma(H^+ \to \tau^+ \nu_\tau)}{\Gamma(H^+ \to \tau^+ \nu_\tau) + \Gamma(H^+ \to c\bar{s})}, \quad (22)$$

where, in the model I (and IV) one has

$$\Gamma(H^+ \to c\bar{s}) = \frac{3g_W^2 m_{H^+}}{32\pi m_W^2} |V_{cs}|^2 (\cot^2 \beta) \lambda^{\frac{1}{2}} (1, \frac{m_c^2}{m_{H^+}^2}, \frac{m_s^2}{m_{H^+}^2})$$
$$\times \left[ (m_c^2 + m_s^2)(1 - \frac{m_c^2}{m_{H^+}^2} - \frac{m_s^2}{m_{H^+}^2}) + 4\frac{m_c^2 m_s^2}{m_{H^+}^2} \right], \quad (23)$$

and for the model II (and III) one has

$$\Gamma(H^+ \to c\bar{s}) = \frac{3g_W^2 m_{H^+}}{32\pi m_W^2} |V_{cs}|^2 \lambda^{\frac{1}{2}} (1, \frac{m_c^2}{m_{H^+}^2}, \frac{m_s^2}{m_{H^+}^2})$$
$$\times \left[ (m_c^2 \cot^2 \beta + m_s^2 \tan^2 \beta)(1 - \frac{m_c^2}{m_{H^+}^2} - \frac{m_s^2}{m_{H^+}^2}) \right.$$
$$\left. - 4\frac{m_c^2 m_s^2}{m_{H^+}^2} \right]. \quad (24)$$

These results are in complete agreement with Ref. [18].

In the limit of $m_i^2/m_H^2 \to 0$ ($i = c, s$), the branching ratio (22) in the type-I 2HDM reads

$$B_\tau^H = \frac{1}{1 + 3|V_{cs}|^2 [(\frac{m_s}{m_\tau})^2 + (\frac{m_c}{m_\tau})^2]}, \quad (25)$$

which is independent of the $\tan \beta$-values and in the type-II 2HDM (MSSM), one has

$$B_\tau^H = \frac{1}{1 + 3|V_{cs}|^2 [(\frac{m_s}{m_\tau})^2 + (\frac{m_c}{m_\tau})^2 \cot^4 \beta]}. \quad (26)$$

Taking $m_c = 1.67$ GeV, $m_\tau = 1.776$ GeV and $|V_{cs}| = 0.9734$, the branching ratio in the model I is $B_\tau^H = 0.284$. In the MSSM scenario, the branching ratio depends on the $\tan \beta$. The dependence of this branching ratio on the $\tan \beta$ is plotted in Fig. 2. As is seen, for the numerical values of $\tan \beta > 5$ that we apply in this work the branching ratio is $B_\tau^H = 1$, to a very high accuracy.

The model-independent bounds on the branching ratio of a light charged Higgs boson are transformed into limits in the $m_{H^\pm} - \tan \beta$ parameter space. Direct searches at the LHC, with a center-of-mass energy of 7 TeV [19–21] and 8 TeV [22,23] set stringent constraints on the parameter space. We will discuss on these constraints when our numerical analysis is presented.

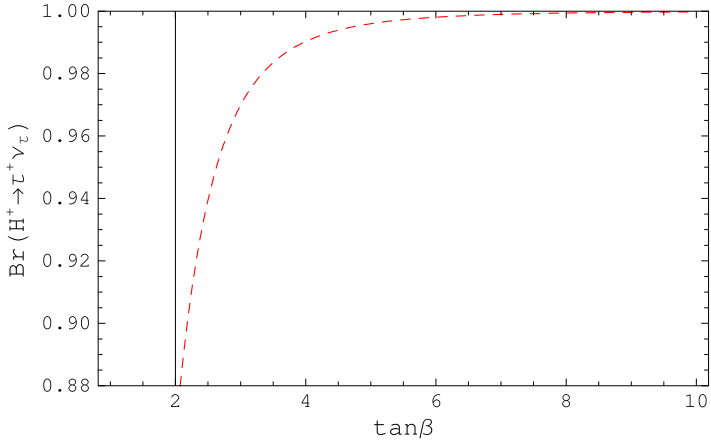

Fig. 2. Lowest order branching ratio for the decay $H^+ \to \tau^+ \nu_\tau$ as a function of $\tan\beta$ in the MSSM scenario.

## 3. Polarized top quark decay in the narrow width approximation

Here, we concentrate on the light charged Higgs inclusively produced through polarized top quark decays $t(\uparrow) \to b + H^+(\to \tau^+\nu_\tau)$. Since, bottom quarks hadronize before they decay, therefore, the study of energy distribution of produced b-jets (bottom- or charmed-flavored mesons in this work) in the following process

$$t(\uparrow) \to b(\to B/D + X) + H^+(\to \tau^+\nu_\tau), \tag{27}$$

is proposed as a new channel to indirect search for the charged Higgs bosons at the LHC. Therefore, of particular interest are to evaluate the distribution in the scaled-energy $(x_B, x_D)$ of B/D-mesons in the top quark rest frame as realistically and reliably as possible. For this study, one needs to evaluate the quantity $d\Gamma/dx_B$ (or $d\Gamma/dx_D$) where, following the notation introduced in [5], the mesonic scaled-energy fraction is defined as

$$x_B = \frac{E_B}{E_b^{max}} \text{ (or } x_D = \frac{E_D}{E_b^{max}}), \tag{28}$$

where $E_b^{max} = (m_t^2 + m_b^2 - m_{H^+}^2)/(2m_t)$. Considering the notation defined in (16), this scaled-energy is simplified as $x_B = E_B/(m_t S)$ (and also $x_D = E_D/(m_t S)$).

To evaluate the partial decay width of process (27) differential in $x_B$ (or correspondingly in $x_D$), we apply the factorization theorem of the QCD-improved parton model [24]. According to this theorem, the energy distribution of B-hadrons might be expressed as the convolution of the parton-level spectrum with the nonperturbative fragmentation function $D_{a=b,g}^B(z, \mu_F)$, describing the hadronization $b/g \to B$,

$$\frac{d\Gamma}{dx_B}(t(\uparrow) \to B(\tau^+\nu_\tau) + X) =$$

$$\sum_{a=b,g} \int_{x_a^{min}}^{x_a^{max}} \frac{dx_a}{x_a} \frac{d\Gamma}{dx_a}(\mu_R, \mu_F) D_a^B(\frac{x_B}{x_a}, \mu_F), \tag{29}$$

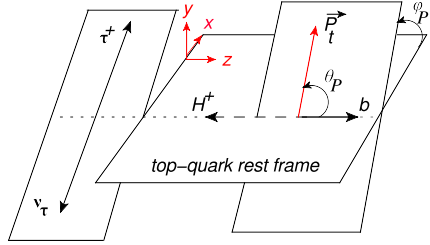

Fig. 3. The azimuthal and polar angles in the helicity coordinate system. $\vec{P}_t$ is the polarization vector of the top quark.

where the $D_g^B(z, \mu_F)$-contribution is appeared at NLO. In (29), $d\Gamma/dx_a$ is the parton-level differential width of the process (27). As in (28), we define the normalized energy fraction of partons (gluon and bottom quark) as $x_a = E_a/E_a^{max} = E_a/(m_t S)$ so that $\eta \leq x_b \leq 1$ and $0 \leq x_g \leq 1$. In (29), $\mu_F$ and $\mu_R$ are the factorization and the renormalization scales, respectively, and generally one can use two different values for these scales. However, a choice often made consists of setting $\mu_R = \mu_F$ and we use the convention $\mu_R = \mu_F = m_t$ in this work.

To achieve the spin-dependent energy distribution of hadrons, at first, we need to calculate the triply differential decay width $d\Gamma/(dx_b d\cos\theta_P d\phi_P)$ of the partonic process $t(\uparrow) \to bH^+ \to b(\tau^+ \nu_\tau)$. Polar and azimuthal angles $\theta_P$ and $\phi_P$ show the orientation of the plane including the spin of the top quark relative to the event plane, see Fig. 3.

Following the narrow width approximation (14), one has

$$\frac{d\Gamma}{dx_b d\cos\theta_P d\phi_P}(t(\uparrow) \to b\tau^+ \nu_\tau) = \frac{d\Gamma}{dx_b d\cos\theta_P d\phi_P}(t(\uparrow) \to bH^+) \times B_\tau^H(H^+ \to \tau^+ \nu_\tau), \quad (30)$$

where in the MSSM, we assume $B_\tau^H(H^+ \to \tau^+ \nu_\tau) = 1$ for $\tan\beta > 5$ (Fig. 2).

For the decay process $t(\uparrow) \to bH^+$, to clarify the correlations between the top quark decay products and the spin of top quark, the general angular distribution of the triply differential decay width $d\Gamma/dx_b$ is given by [25]

$$\frac{d\Gamma}{dx_b d\cos\theta_P d\phi_P} = \frac{1}{4\pi}\left\{\frac{d\Gamma_U}{dx_b} + P\frac{d\Gamma_\theta}{dx_b}\cos\theta_P + P\frac{d\Gamma_\phi}{dx_b}\sin\theta_P \cos\phi_P\right\}, \quad (31)$$

where, $P$ is the magnitude of the top quark polarization. $P = 0$ is for an unpolarized top quark while $P = 1$ stands for a 100% polarized top quark. In (31), $d\Gamma_U/dx_b$ corresponds to the unpolarized differential decay rate while $d\Gamma_\theta/dx_b$ and $d\Gamma_\phi/dx_b$ describe the polar and azimuthal correlations between the polarization of the top quark and its decay products, respectively.

For the analysis of spin-momentum correlations between the top polarization vector and the momenta of its decay products, we consider the helicity frame shown in Fig. 3. In this frame, the three-momentum of b-quark is defined as $\vec{p}_b \| (+\hat{z})$, and the polarization vector of the top quark is evaluated with respect to the positive $\hat{z}$-axis.

Concerning the importance of this helicity frame, we take a moment to explain it in detail. Ignoring the interference terms, in the NWA the total decay width of an unpolarized top quark is given by $\Gamma^{tot} = \Gamma^{SM}_{t \to bW^+} + \Gamma^{MSSM}_{t \to bH^+}$. The same result is valid for the scaled-energy ($x_B$) distribution of the produced B-meson, i.e. $d\Gamma^{tot}/dx_B = d\Gamma(t \to BW^+)/dx_B + d\Gamma(t \to BH^+)/dx_B$. In other words, to obtain the total energy spectrum of B-mesons in the unpolarized top decays all decay modes including $t \to B + W^+/H^+$ should be summed up. In the presence of a polarized top quark, the same consequence is valid as long as the polarization of top quark is measured relative to the same three-momentum, for example, the momentum of bottom quark in our work. Since, the hadronization mechanism of bottom quarks is the same both in the SM and beyond SM theories, therefore, the study of energy spectrum of observed mesons in the introduced helicity frame can be considered as a new channel to indirect search for the charged Higgs bosons. In fact, any deviation of the total energy spectrum of produced meson from the corresponding SM predictions (presented in [16]) can be associated with the existence of charged Higgs bosons.

Now, we back to Eq. (31). $d\Gamma_U/dx_b$ describes the unpolarized differential decay rate of the top quark which is independent of the selected frame. The $\mathcal{O}(\alpha_s)$ radiative corrections to this rate are extensively studied in our previous work [6]. Here, we present our analytical results for the NLO radiative corrections to the angular correlation functions ($d\Gamma_\theta/dx_b$ and $d\Gamma_\phi/dx_b$) in the helicity frame shown in Fig. 3. Finally, at the hadron level we shall present and compare our predictions for the energy distribution of B/D-mesons, considering all contributions.

### 3.1. Born term results for the polarized top decay

It is straightforward to compute the Born term contribution to the partial decay rate of a polarized top quark in the 2HDM. The Born term amplitude squared for the process $t(\uparrow) \to bH^+$ reads

$$|M_0|^2 = 2(p_b \cdot p_t)(a^2 + b^2) + 2(a^2 - b^2)m_b m_t \\ + 4ab m_t(p_b \cdot s_t), \qquad (32)$$

where we replaced $\sum_{s_t} u(p_t, s_t)\bar{u}(p_t, s_t) = (\not{p}_t + m_t)$ in the unpolarized Dirac string by $u(p_t, s_t)\bar{u}(p_t, s_t) = (1 - \gamma_5 \not{s}_t)(\not{p}_t + m_t)/2$ in the polarized state. This result is converted to the relation (17) if one sets $s_t = 0$.

Considering Fig. 3, the top polarization four-vector is set as $s_t = P(0; \sin\theta_P \cos\phi_P, \sin\theta_P \sin\phi_P, \cos\theta_P)$, whereas one has $p_b \cdot s_t = -P|\vec{p}_b|\cos\theta_P$. Therefore, at the Born level the helicity structure of partial rate of the process $t(\uparrow) \to bH^+ \to b\tau^+ \nu_\tau$ reads

$$\frac{d^2 \tilde{\Gamma}^{(0)}}{d\cos\theta_P d\phi_P} = \frac{B^H_\tau}{4\pi} \left\{ \tilde{\Gamma}^{(0)}_U - P\tilde{\Gamma}^{(0)}_\theta \cos\theta_P \\ + P\tilde{\Gamma}^{(0)}_\phi \sin\theta_P \cos\phi_P \right\}, \qquad (33)$$

where $B^H_\tau(H^+ \to \tau^+ \nu_\tau) \approx 1$ in the MSSM scenario. The unpolarized Born-level rate $\tilde{\Gamma}^{(0)}_U(t \to bH^+)$ is given in (18), and for the polar and azimuthal correlation functions $\tilde{\Gamma}^{(0)}_\theta$ and $\tilde{\Gamma}^{(0)}_\phi$, one has

$$\tilde{\Gamma}^{(0)}_\theta = m_t Q^2 \left(\frac{ab}{2\pi}\right), \\ \tilde{\Gamma}^{(0)}_\phi = 0. \qquad (34)$$

In (34), the product of two coupling factors in the MSSM scenario reads

$$ab = \frac{G_F}{\sqrt{2}}|V_{tb}|^2(m_t^2 \cot^2\beta - m_b^2 \tan^2\beta). \tag{35}$$

Considering Eqs. (18) and (34), it is seen that the polarized and unpolarized decay widths have a minimum at $\tan\beta = \sqrt{m_t/m_b} \approx 6$ if one takes $m_t = 172.9$ GeV and $m_b = 4.78$ GeV. Therefore, the $t \to BH^+$ branching fraction has a pronounced dip at $\tan\beta \approx 6$. Although, this is partly compensated by a large value of the $H^+ \to \tau^+\nu_\tau$ branching fraction, which is $B_\tau^H(H^+ \to \tau^+\nu_\tau) \approx 1$ for $\tan\beta > 5$, the product still has a significant dip at $\tan\beta \approx 6$ [17]. This state is not established for the type-I 2HDM scenario, as Eq. (25) shows that the branching ratio is independent of $\tan\beta$.

The fact that $\tilde{\Gamma}_\phi^{(0)} = 0$ means that the azimuthal correlation measurement has zero analyzing power to analyze the polarization of the top quark and the nonzero contribution arises from the radiative corrections.

### 3.2. NLO contribution to the polar and azimuthal differential decay rates; $d\tilde{\Gamma}_\theta^{NLO}/dx_b$ and $d\tilde{\Gamma}_\phi^{NLO}/dx_b$

Basically, the required ingredients for the NLO QCD perturbative calculations are the virtual one-loop and the real gluon emission contributions. At $\mathcal{O}(\alpha_s)$, therefore, the full amplitude for the decay process $t(\uparrow) \to bH^+$ is the sum of the amplitudes of the Born term $M_0$, the virtual one-loop $M_{\text{loop}}^{(\alpha_s)}$ and the real contributions $M_{\text{real}}^{(\sqrt{\alpha_s})}$, so that the NLO decay amplitude squared reads $|M|^2 = |M_0|^2 + |M_{\text{vir}}|^2 + |M_{\text{real}}|^2 + \mathcal{O}(\alpha_s^2)$ where $|M_{\text{vir}}|^2 = 2Re(M_0^\dagger \cdot M_{\text{loop}})$ and $|M_{\text{real}}|^2 = Re(M_{\text{real}}^\dagger \cdot M_{\text{real}})$.

Since, the contribution of azimuthal correlation to the Born amplitude is zero, then $d\tilde{\Gamma}_\phi^{\text{vir}}/dx_b = 0$. It means, the virtual one-loop corrections are contributed in the unpolarized rate $(d\tilde{\Gamma}_U^{\text{vir}}/dx_b)$ and in the polar correlation function $(d\tilde{\Gamma}_\theta^{\text{vir}}/dx_b)$. Thus, the $\mathcal{O}(\alpha_s)$ radiative corrections to the $\tilde{\Gamma}_\phi$ (and $d\tilde{\Gamma}_\phi/dx_b$) just result from the real gluon emissions. Due to the soft treatments of the real gluons emitted, the corresponding Feynman diagrams include the infrared (IR) singularities. Generally, to extract the ultraviolet (UV) singularities (which appear in the virtual corrections) and the IR divergences we work in D-dimensional regularization scheme where $D \neq 4$. In this scheme, as an example, the differential decay rate for the real corrections to the $tbH^+$-vertex is given by

$$d\tilde{\Gamma}^{\text{real}} = \frac{\mu_F^{2(4-D)}}{2m_t}|M^{\text{real}}|^2 dR_3(p_t, p_b, p_g, p_{H^+}), \tag{36}$$

where, $dR_3$ stands for the 3-body phase space element including the three-momenta of outgoing particles, i.e. $\vec{p}_b, \vec{p}_g$ and $\vec{p}_{H^+}$. Here $M^{\text{real}}$ is

$$M^{\text{real}} = g_s \frac{\lambda^a}{2}\bar{u}(p_b, s_b)\left\{\frac{2p_t^\mu - \slashed{p}_g\gamma^\mu}{2p_t \cdot p_g}\right. \tag{37}$$

$$\left. - \frac{2p_b^\mu + \slashed{p}_g\gamma^\mu}{2p_b \cdot p_g}\right\}(a\mathbf{1} + b\gamma_5)u(p_t, s_t)\epsilon_\mu^\star(p_g, s_g),$$

where the polarization vector of the real gluon is denoted by $\epsilon(p_g, s_g)$. The first and second terms in the curly brackets refer to the real gluon emissions from the top and bottom quarks, respectively.

In the dimensional regularization approach, the angular integral is written as

$$\frac{d\Omega}{d\phi_P d\cos\theta_P} = -\frac{\pi^{\frac{D-3}{2}}}{\Gamma(\frac{D-3}{2})}(\sin\theta_P)^{D-4}(\sin\phi_P)^{D-4}. \tag{38}$$

More detail about this approach can be found in our previous work [16].

As a subtle issue of dealing with $\gamma^5 = i\gamma^0\gamma^1\gamma^2\gamma^3$ in dimensional regularization (DR) scheme, note that, $\gamma^5$ is not well defined in D-dimensions and there is no unique way to handle $\gamma^5$ in DR. The anticommutation relation $\{\gamma^5, \gamma^\mu\} = 0$ produces ambiguity so one can not simply apply this relation in general D-dimensions. In other words, the above anticommutation relation along with the relation $Tr[\gamma^5\gamma^\alpha\gamma^\beta\gamma^\gamma\gamma^\delta] \neq 0$ can not be simultaneously satisfied in DR scheme. There are several prescriptions to prevent this ambiguity, see for example [26–29]. In this work, we employ the Breitenlohner–Maison–t'Hooft–Veltman scheme [29,30] in which we accept that $\gamma^5$ is a purely 4-dimensional object and therefore does not anitcommute with D-dimensional Dirac matrices, i.e. $\{\gamma^5, \gamma^\mu\} \neq 0$, while the relation $Tr[\gamma^5\gamma^\alpha\gamma^\beta\gamma^\gamma\gamma^\delta] \neq 0$ still holds in D-dimensions.

Since the lowest-order term contribution to $\tilde{\Gamma}_\phi$ vanishes (34), then one does not need to introduce any IR regularization scheme such as a fictitious gluon mass or dimensional regularization when calculating the azimuthal correlation. In other words, the tree-graph contributions to the $\tilde{\Gamma}_\phi$ are IR-finite at NLO.

To calculate the azimuthal differential rate $d\tilde{\Gamma}_\phi/dx_b$, in (36) by working in four dimensions we fix the 3-momentum of the b-quark and integrate over the gluon energy, which ranges as

$$\frac{m_t S(1-x_b)}{1+R-2Sx_b}F_- \leq E_g \leq \frac{m_t S(1-x_b)}{1+R-2Sx_b}F_+, \tag{39}$$

where $F_\pm = 1 - Sx_b \pm S\sqrt{x_b^2 - \eta^2}$. Therefore, one has

$$\frac{d\tilde{\Gamma}_\phi}{dx_b} = \tilde{\Gamma}_\theta^{(0)}\frac{\alpha_s(\mu_R)C_F}{2Q^2}\frac{S+\sqrt{R}}{\sqrt{x_b^2-\eta^2}}(Sx_b + \sqrt{R}), \tag{40}$$

where $C_F = 4/3$ stands for the color factor and $\alpha_s(\mu_R)$ is the strong coupling constant. Other parameters are defined in (16).

By integrating over $x_b$, so $\eta \leq x_b \leq 1$, one has

$$\tilde{\Gamma}_\phi = \tilde{\Gamma}_\theta^{(0)}\frac{\alpha_s C_F}{2Q^2}(S+\sqrt{R})(Q + \sqrt{R}\ln\frac{S+Q}{\sqrt{R}}). \tag{41}$$

In the massless scheme where the $m_b = 0$ approximation is considered, this result simplifies as $\hat{\Gamma}_\phi = \hat{\Gamma}_\theta^{(0)}C_F\alpha_s/2$.

Since the observed mesons in top decays can be also produced through a fragmenting real gluon, therefore, to obtain the most accurate energy spectrum of mesons one has to add the contribution of gluon fragmentation to the b-quark one. Then, one also needs the azimuthal partial decay rate $d\hat{\Gamma}_\phi/dx_g$, where $x_g = E_g/E_b^{max}$ is the scaled-energy fraction of the real gluon, as in (28). In [5,6], we showed that the gluon contribution would be important at a low energy of the detected meson so this contribution decreases the size of decay rate at the threshold energy. In the helicity frame selected, the azimuthal differential decay width $d\hat{\Gamma}_\phi/dx_g$ is given by

$$\frac{d\hat{\Gamma}_\phi}{dx_g} = \hat{\Gamma}_\theta^{(0)}\frac{\alpha_s C_F}{2S}\frac{1-Sx_g(1+x_g)-\sqrt{1-2Sx_g}}{x_g^2\sqrt{1-2Sx_g}}. \tag{42}$$

In [5], we showed that the $g \to B$ contribution into the NLO energy spectrum of B-meson is negative and appreciable only in the low-$x_B$ region and for higher values of $x_B$ the NLO result is practically exhausted by the $b \to B$ contribution. The contribution of gluon is calculated to see where it contributes to $d\Gamma/dx_B$ and can not be discriminated as an experimental quantity.

The NLO expression for the $d\tilde{\Gamma}_\theta/dx_b$ is obtained by summing the Born term, the virtual one-loop and the real gluon contributions. To extract all IR- and UV-singularities one needs to work in D-dimensions considering the relation (38).

The QCD virtual one-loop corrections arise from emission and absorption of a virtual gluon from the same quark leg (quark self-energy) and from a virtual gluon exchanged between the top and bottom quark legs (vertex correction). The self-energy Feynman diagrams are related to the mass and wave function renormalizations of both top and bottom quarks, more detail can be found in [31]. Adopting the on-shell mass-renormalization scheme, the virtual one-loop corrections include both the IR- and UV-divergences so the UV ones appear when the integration region of the internal momentum of the virtual gluon goes to infinity and the IR-divergences arise from the soft-gluon singularities. The UV-divergences are canceled after summing all virtual corrections up but the IR-singularities are remaining. To remove the remaining IR-divergences, one needs to consider the real gluon corrections at NLO.

The NLO real contributions result from the real gluon emissions from the bottom and top quarks, individually. In calculation of the real gluon corrections, since we preserve the mass of b-quark from the beginning, there are no collinear singularities and all IR-divergences arise from the soft real gluon emission. According to the Lee-Nauenberg theorem, all singularities cancel each other after summing all contributions up, and the final result is free of IR singularities. More detail about the NLO corrections and the D-dimensional regularization scheme can be found in [6] where we calculated the unpolarized differential decay rate $d\tilde{\Gamma}_U/dx_i$ ($i = b, g$). The technical detail of calculations are the same.

Ignoring the detail of our calculations, the NLO expression for the $d\tilde{\Gamma}_\theta/dx_b$ is as follows

$$\frac{d\tilde{\Gamma}_\theta}{dx_b} = \tilde{\Gamma}_\theta^{(0)} \Bigg\{ \delta(1-x_b) + \frac{C_F \alpha_s(\mu_R)}{2\pi Q} \Bigg\{$$

$$\delta(1-x_b) \Bigg[ -2Q(1 + 2\ln\frac{2S(1-\eta)}{\sqrt{y}}) +$$

$$4S[Li_2(S+Q) - Li_2(S-Q) - Li_2(\frac{2Q}{S+Q})]$$

$$+(R-1)\ln\frac{1-S-Q}{1+Q-S} - \ln R \Big(\frac{3(a^2+b^2)Q}{4ab}$$

$$+\frac{2Q(S-1)}{y} + S\ln\frac{1+Q-S}{1-S-Q}\Big) -$$

$$4S\ln\frac{S-Q}{\sqrt{R}} \Big( \ln(2S(1-\eta)) + \frac{1}{2}\ln\frac{y}{R} +$$

$$\ln\frac{S-Q}{\sqrt{R}} + \frac{S(1+y) + R(S-y-2)}{2Sy} \Big) \Bigg]$$

$$-\frac{2TS^2}{Q\sqrt{x_b^2-\eta^2}} \Big(1 + x_b + \frac{2(x_b^2-\eta^2)}{(1-x_b)_+}\Big) \Bigg\} \Bigg\}, \tag{43}$$

where the plus distribution is defined as usual, and

$$T = \sqrt{x_b^2 - \eta^2} - x_b \ln \frac{\eta}{x_b - \sqrt{x_b^2 - \eta^2}}. \tag{44}$$

Also, the gluon contribution is expressed as

$$\begin{aligned}\frac{d\tilde{\Gamma}_\theta}{dx_g} = \tilde{\Gamma}_\theta^{(0)} \frac{\alpha_s(\mu_R)}{2\pi} C_F \Bigg\{ & 2R + 3 - \frac{R + y^2}{4S(1 - 2Sx_g)^2} + \\ & \frac{1}{Sx_g^2} \ln(1 + R - 2Sx_g) + \frac{(1 - 4S)^2 - 4S^2 + R}{4S(1 - 2Sx_g)} \\ & - \frac{1 + (1 - x_g^2)}{x_g}(1 + 2\ln x_g + \ln R) - \frac{x_g}{2} + \\ & \frac{1 + (1 - x_g)^2}{x_g} \ln \frac{4S^2 x_g^2 (1 - x_g)^2}{1 - 2Sx_g} \Bigg\}. \end{aligned} \tag{45}$$

The analytical results presented in (40)–(45) are completely new and are being presented for the first time. Specifically, the contribution of azimuthal correlation $\Gamma_\phi$ has not been already calculated. One needs these angular differential decay rates to obtain the energy spectrum of produced mesons in polarized top decays, see (29).

## 4. GM-VFN scheme

As was mentioned, our main aim is to evaluate the scaled-energy distribution of the $B/D$-mesons produced in the inclusive process $t(\uparrow) \to H^+ B/D + X$ followed by $H^+ \to \tau^+ \nu_\tau$, where $X$ stands for the unobserved final state. Therefore, we calculate the NLO decay width of the corresponding process differential in $x_B$ ($d\Gamma/dx_B$) and $x_D$ ($d\Gamma/dx_D$) in the general-mass variable-flavor-number scheme (GM-VFNs), where $x_B$ and $x_D$ are defined in (28). In the top quark rest frame applied in our work, the B-meson has the energy $E_B = p_t \cdot p_B/m_t$, where $m_B \leq E_B \leq [m_t^2 + m_B^2 - m_{H^+}^2]/(2m_t)$. In the case of gluon fragmentation, $g \to B$, it has energy $m_B \leq E_B \leq [m_t^2 + m_B^2 - (m_b + m_{H^+})^2]/(2m_t)$ [5]. The same results hold for the D-meson with the mass $m_D$.

Considering the factorization theorem (29), the energy spectrum of B/D-meson can be obtained by the convolution of the convenient parton-level spectrum with the nonperturbative fragmentation function $D_{i=b,g}^B(z, \mu_F)$ (or $D_{i=b,g}^D(z, \mu_F)$). We will discuss about these FFs in section 5.

Now, we explain how the quantity $d\Gamma_\theta(\mu_R, \mu_F)/dx_i$ (with $i = b, g$) will have to be evaluated in the GM-VFN scheme. In Sec. 3.2, we employed the Fixed-Flavor-Number (FFN) scheme which contains the full $m_b$ dependence. In the FFN scheme, the large logarithmic singularities of the form $(\alpha_s/\pi) \ln(m_t^2/m_b^2)$ spoil the convergence of the perturbative expansion when $m_b/m_t \to 0$. The massive or GM-VFN scheme is devised to resume these large logarithms and to retain the whole nonlogarithmic $m_b$ dependence at the same time and this is achieved by introducing appropriate subtraction terms in the NLO FFN expression for $d\tilde{\Gamma}_\theta(\mu_R, \mu_F)/dx_i$. In this case, the NLO ZM-VFN results are exactly recovered in the limit $m_b/m_t \to 0$. In the GM-VFN scheme, the subtraction terms are constructed as

$$\frac{1}{\Gamma_\theta^{(0)}} \frac{d\Gamma_\theta^{\mathbf{Sub}}}{dx_i} = \lim_{m_b \to 0} \frac{1}{\tilde{\Gamma}_\theta^{(0)}} \frac{d\tilde{\Gamma}_\theta^{\mathbf{FFN}}}{dx_i} - \frac{1}{\hat{\Gamma}_\theta^{(0)}} \frac{d\hat{\Gamma}_\theta^{\mathbf{ZM}}}{dx_i}, \tag{46}$$

where $1/\tilde{\Gamma}_\theta^{(0)} \times d\tilde{\Gamma}_\theta^{\mathbf{FFN}}/dx_i$ are given in (43) and (45) and $1/\hat{\Gamma}_\theta^{(0)} \times d\hat{\Gamma}_\theta^{\mathbf{ZM}}/dx_i$ are the partial decay rates computed in the ZM-VFN scheme [7], in which all information on the $m_b$-dependence is wasted.

In conclusion, the GM-VFN results are obtained by subtracting the subtraction terms from the FFN ones [32,33], i.e.

$$\frac{1}{\Gamma_\theta^{(0)}} \frac{d\Gamma_\theta^{\mathbf{GM}}}{dx_i} = \frac{1}{\tilde{\Gamma}_\theta^{(0)}} \frac{d\tilde{\Gamma}_\theta^{\mathbf{FFN}}}{dx_i} - \frac{1}{\Gamma_\theta^{(0)}} \frac{d\Gamma_\theta^{\mathbf{Sub}}}{dx_i}. \tag{47}$$

Taking the limit $m_b \to 0$ in Eqs. (43) and (45), one obtains the following subtraction terms

$$\frac{1}{\Gamma_\theta^{(0)}} \frac{d\Gamma_\theta^{\mathbf{Sub}}}{dx_b} = \frac{\alpha_s(\mu_R)}{2\pi} C_F \times \left\{ \frac{1 + x_b^2}{1 - x_b} \left[ \ln \frac{\mu_F^2}{m_b^2} - 2\ln(1 - x_b) - 1 \right] \right\}_+, \tag{48}$$

and

$$\frac{1}{\Gamma_\theta^{(0)}} \frac{d\Gamma_\theta^{\mathbf{Sub}}}{dx_g} = \frac{\alpha_s(\mu_R)}{2\pi} C_F \frac{1 + (1 - x_g)^2}{x_g} \times \left( \ln \frac{\mu_F^2}{m_b^2} - 2\ln x_g - 1 \right). \tag{49}$$

The result (48) coincides with the perturbative FF of the transition $b \to b$ [34–41] and is in consistency with the Collin's factorization theorem [24], which guarantees that the subtraction terms are universal. Thus, the result obtained in (48) ensures the correctness of our result shown in (43). In [5,42], the GM-VFN scheme is applied to study the NLO energy spectrum of B-meson in the process $t \to bW^+$. There was shown, Eq. (48) coincides with the perturbative FF of the transition $b \to b$ as well.

## 5. Numerical analysis

In the MSSM, the charged Higgs mass is restricted by $m_{H^\pm} > m_{W^\pm}$ at tree-level [43], however, this restriction does not hold for some regions of the MSSM $\tan\beta - m_{H^\pm}$ parameter space after including radiative corrections. In Ref. [11], it is mentioned that a charged Higgs having a mass in the range $80 \text{ GeV} \leq m_{H^\pm} \leq 160 \text{ GeV}$ is a logical possibility and its effects should be searched for in the decay chain $t \to bH^+(\to \tau^+ \nu_\tau)$. Searches for charged Higgs bosons have already been started at the Tevatron and, at the present, the last results in 19.5–19.7 fb$^{-1}$ of proton-proton collision data recorded at $\sqrt{s} = 8$ TeV are reported by the CMS [44] and the ATLAS [45] collaborations, using the $\tau + jets$ channel with a hadronically decaying $\tau$ lepton in the final state. According to these results, the large region in the $(m_{H^+} - \tan\beta)$-plane is excluded for $m_{H^+} = 80$–160 GeV and the only unexcluded regions of this parameter space include the charged Higgs masses as $90 \leq m_{H^+} \leq 100$ GeV (for $6 < \tan\beta < 10$) and $140 \leq m_{H^+} \leq 160$ GeV (for $3 < \tan\beta < 21$). These regions along with the $\pm 1\sigma$ band around the expected limit are shown in Fig. 4 which is taken from Ref. [45]. A same exclusion is reported by the CMS collaboration [44]. In this work, we restrict ourselves to these unexcluded regions. Although, a definitive search over these remaining parts of the $m_{H^+} - \tan\beta$ parameter space still has to be carried out by the LHC experiments.

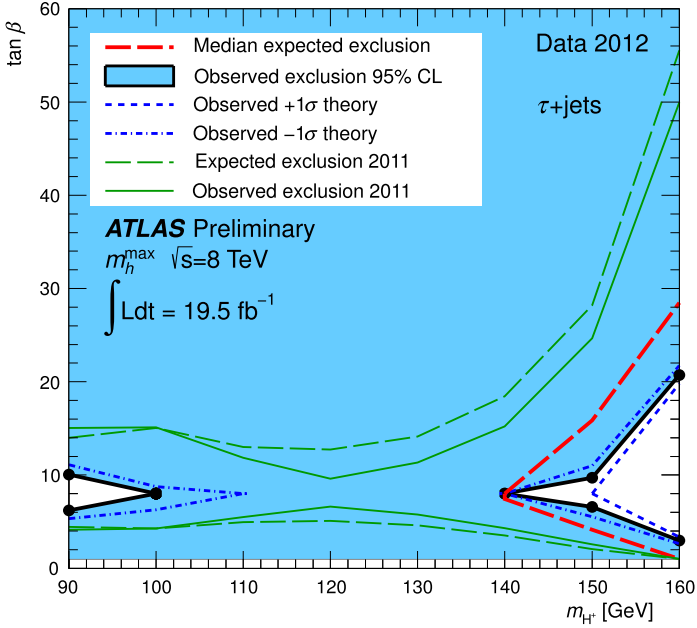

Fig. 4. Exclusion region in the MSSM $\tan\beta - m_{H^+}$ parameter space for $m_{H^+} = 80 - 160$ GeV is shown. The $\pm 1\sigma$ band around the expected limit is also shown [45].

To present our numerical analysis, we adopt the input parameter values from Ref. [46] as; $G_F = 1.16637 \times 10^{-5}$ GeV$^{-2}$, $m_t = 172.98$ GeV, $m_b = 4.78$ GeV, $m_B = 5.279$ GeV, $m_D = 1.87$ GeV, $m_W = 80.399$ GeV.

Having the differential decay rates at the parton level, in the first step we turn to our numerical predictions of polarized and unpolarized decay rates, while the strong coupling constant is evolved from $\alpha_s(m_Z) = 0.1184$ to $\alpha_s(m_t) = 0.1070$. Considering our analytical results for the unpolarized decay width $\Gamma_U$ [6], the azimuthal correlation function $\tilde{\Gamma}_\phi$ (41) and the polar correlation rate $\tilde{\Gamma}_\theta$ (43), and by taking $m_{H^+} = 160$ GeV and $\tan\beta = 16$ from Fig. 4, in the type-I 2HDM one has

$$\frac{d^2\Gamma_{\text{NLO}}^{\text{Model I}}}{d\phi_P d\cos\theta_P}(t \to b\tau^+\nu_\tau) = \frac{B_\tau^H}{4\pi}\left\{\Gamma_U^{(0)}(1 - 0.0058)\right.$$
$$- \tilde{\Gamma}_\theta^{(0)}(1 - 0.026)P\cos\theta_P$$
$$\left. + \tilde{\Gamma}_\theta^{(0)}(0.1775)P\sin\theta_P\cos\phi_P\right\}$$
$$= B_\tau^H \times \frac{\Gamma}{4\pi}\left\{1 - (0.923)P\cos\theta_P\right.$$
$$\left. + (0.168)P\sin\theta_P\cos\phi_P\right\}, \qquad (50)$$

where $\Gamma = 125 \times 10^{-6}$ GeV and $B_\tau^H(H^+ \to \tau^+ \nu_\tau) = 0.284$. The corresponding result in the type-II 2HDM (MSSM) scenario reads

$$\frac{d^2\Gamma_{\text{NLO}}^{\text{Model II}}}{d\phi_P d\cos\theta_P}(t \to b\tau^+\nu_\tau) = \frac{B_\tau^H}{4\pi}\Big\{\Gamma_U^{(0)}(1 - 0.465)$$
$$- \tilde{\Gamma}_\theta^{(0)}(1 - 0.525)P\cos\theta_P$$
$$+ \tilde{\Gamma}_\theta^{(0)}(0.177)P\sin\theta_P\cos\phi_P\Big\}$$
$$= B_\tau^H \times \frac{\Gamma}{4\pi}\Big\{1 - (0.260)P\cos\theta_P$$
$$+ (0.097)P\sin\theta_P\cos\phi_P\Big\}, \tag{51}$$

where $\Gamma = 106 \times 10^{-4}$ GeV and $B_\tau^H \approx 1$.

Taking $m_{H^+} = 95$ GeV and $\tan\beta = 8$ from Fig. 4, in the type-I 2HDM scenario, one has

$$\frac{d^2\Gamma_{\text{NLO}}^{\text{Model I}}}{d\phi_P d\cos\theta_P}(t \to b\tau^+\nu_\tau) = \frac{B_\tau^H}{4\pi}\Big\{\Gamma_U^{(0)}(1 - 0.0846)$$
$$- \tilde{\Gamma}_\theta^{(0)}(1 - 0.118)P\cos\theta_P$$
$$+ \tilde{\Gamma}_\theta^{(0)}(0.096)P\sin\theta_P\cos\phi_P\Big\}$$
$$= B_\tau^H \times \frac{\Gamma}{4\pi}\Big\{1 - (0.812)P\cos\theta_P$$
$$+ (0.089)P\sin\theta_P\cos\phi_P\Big\}, \tag{52}$$

where $\Gamma = 139 \times 10^{-4}$ GeV and $B_\tau^H = 0.284$, as before. The corresponding result in the type-II 2HDM scenario is

$$\frac{d^2\Gamma_{\text{NLO}}^{\text{Model II}}}{d\phi_P d\cos\theta_P}(t \to b\tau^+\nu_\tau) = \frac{B_\tau^H}{4\pi}\Big\{\Gamma_U^{(0)}(1 - 0.460)$$
$$- \tilde{\Gamma}_\theta^{(0)}(1 - 0.837)P\cos\theta_P$$
$$+ \tilde{\Gamma}_\theta^{(0)}(0.097)P\sin\theta_P\cos\phi_P\Big\}$$
$$= B_\tau^H \times \frac{\Gamma}{4\pi}\Big\{1 - (0.053)P\cos\theta_P$$
$$+ (0.032)P\sin\theta_P\cos\phi_P\Big\}, \tag{53}$$

where $\Gamma = 830 \times 10^{-4}$ GeV and $B_\tau^H \approx 1$.

As is seen, the NLO radiative corrections to the polarized and unpolarized rates are significant, specifically, when the type-II 2HDM scenario is concerned. Also, the azimuthal decay rates are quite small in both models, as is expected, so it seems that their measurements are difficult. Since highly polarized top quarks with more accuracy will become available at higher luminosity hadron colliders through single top production processes [8], then it might then be feasible to experimentally measure these azimuthal correlations.

Apart from the above results, in the following we present our phenomenological predictions for the scaled-energy ($x_B$) distribution of bottom-flavored hadrons and also the $x_D$-distribution of charmed-flavored hadrons inclusively produced in polarized top quark decays and compare them with the unpolarized ones. From now on, we restrict ourselves to the type-II 2HDM (MSSM) scenario. The same study can be done for the type-I 2HDM scenario just by considering the corresponding model parameters presented in (5). Applying the GM-VFN scheme [31] and by considering the factorization theorem (29) we present our results for the scaled-energy spectrum of B/D-mesons in the MSSM. For our study, we consider the quantities $d\Gamma(t(\uparrow) \to B\tau^+\nu_\tau + X)/dx_B$ and $d\Gamma(t(\uparrow) \to D\tau^+\nu_\tau + X)/dx_D$.

Table 1
Values of fit parameters for $b \to D^0/D^+$ FFs at the starting scale $\mu_0 = m_b$ along with the values of $\overline{\chi^2}$ achieved.

|       | N    | a    | γ    | $\overline{\chi^2}$ |
|-------|------|------|------|---------------------|
| $D^0$ | 80.8 | 5.77 | 1.15 | 4.66                |
| $D^+$ | 163  | 6.93 | 1.40 | 2.21                |

In the factorization formula for determination of $x_B$-distribution, the function $D_b^B(z, \mu_F)$ (or $D_b^D(z, \mu_F)$ when $x_D$-distribution is concerned) is the nonperturbative fragmentation function (FF) describing the splitting of $b \to B$ (or $b \to D$ in the $D_b^D(z, \mu_F)$-FF) at the desired scale $\mu_F$. Considering the description presented in Sec. 3.2 about the importance of the gluon splitting into the hadrons, one also needs the $g \to B/D$ FFs. These nonperturbative FFs are universal and process independent. Several models have been yet proposed to describe these hadronization processes. In [47], the FFs of $D^0$- and $D^+$-mesons are determined at NLO by fitting the experimental data from the BELLE, CLEO, OPAL, and ALEPH collaborations in the modified minimal-subtraction ($\overline{MS}$) factorization scheme. There, authors have parametrized the FFs of $b \to D^0/D^+$ at the starting scale $\mu_0 = m_b$, as

$$D_b^{D^0/D^+}(z, \mu_0) = Nz^{-(1+\gamma^2)}(1-z)^a e^{-\gamma^2/z}, \tag{54}$$

while the FF of gluon is set to zero. This parametrization is known as Bowler model. These FFs are evolved to higher scales using the DGLAP evaluation equations [48]. The values of fit parameters together with the achieved values of $\overline{\chi^2}$ are presented in Table 1.

For the $b \to B$ hadronization process, from Ref. [49] we employ the FFs determined at NLO through a global fit to $e^+e^-$-annihilation data taken by ALEPH and OPAL at CERN LEP1 and by SLD at SLAC SLC. In [49], a power model as $D_b^B(z, \mu_0) = Nz^\alpha(1-z)^\gamma$ is used at the initial scale $\mu_0 = 4.5$ GeV, while the gluon FF is generated via the DGLAP evolution. The fit yielded $N = 4684.1$, $\alpha = 16.87$, and $\gamma = 2.628$ with $\overline{\chi^2} = 1.495$.

Considering the unexcluded MSSM $m_{H^+} - \tan\beta$ parameter space shown in Fig. 4, in Fig. 5 we present the $x_B$ distribution of $d\Gamma/dx_B$ considering the azimuthal (dotted line) and the polar (solid line) correlation contributions. Here, we set $\tan\beta = 8$ and $m_H = 95$ GeV. For a more quantitative comparison, the unpolarized contribution (dashed line) is also shown. Here, the B-hadron mass creates a threshold at $x_B = m_B/(m_t S) \approx 0.087$. As was explained in section 3.1, the azimuthal correlation rate is zero at LO, which explains the smallness of the corresponding result at NLO.

In Fig. 6, the same analysis is presented but taking $m_H = 150$ GeV; a charged Higgs mass which is not excluded in the MSSM $m_{H^+} - \tan\beta$ parameter space. Here, the B-meson mass creates a threshold at $x_B \approx 0.22$. From these two plots it is seen that in the unpolarized top decay the partial decay width at the hadron level is always higher than the one in the polarized top decay, specially, in the peak region.

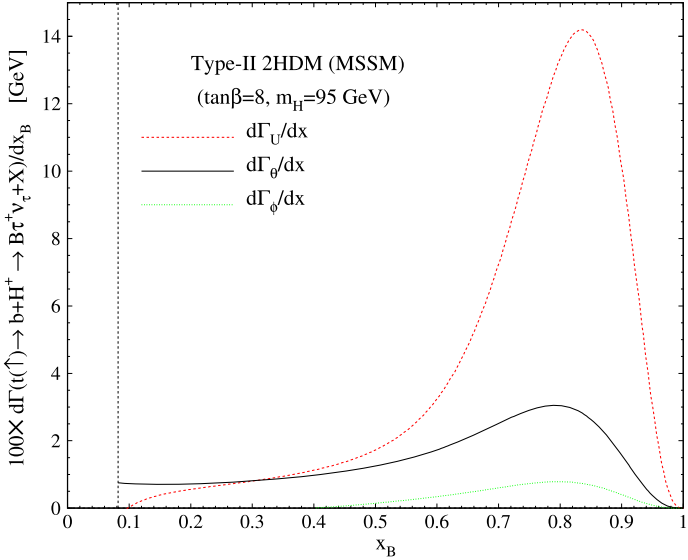

Fig. 5. $x_B$ distribution of $d\Gamma^{NLO}/dx_B$ in the MSSM scenario, considering the unpolarized (dashed line), the polar (solid line) and the azimuthal (dotted line) contributions. Threshold is also shown.

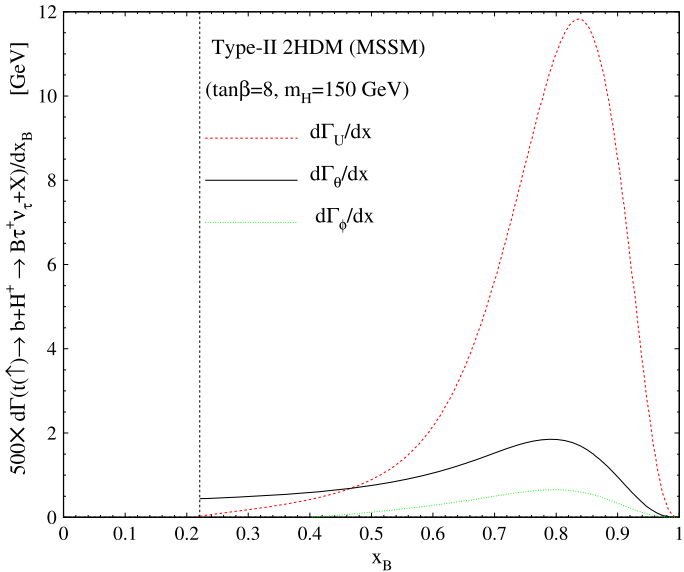

Fig. 6. As in Fig. 5 but for $m_{H^+} = 150$ GeV. Threshold is at $x_B = 0.22$.

In Figs. 7 and 8, the same comparisons are done for the transitions $(b, g) \to D^+$ and $(b, g) \to D^0$, respectively. These figures show that the chance to produce the charmed-flavored mesons through top quark decays is zero in the high-$x_D$ range ($x_D \geq 0.7$). Furthermore, the peak position is shifted towards lower values of the scaled-energy in comparison with the bottom-flavored hadrons. Here, the threshold occurs at $x_D = max\{m_D/(m_t S), \eta\} \approx 0.08$, due to the term $\sqrt{x_b^2 - \eta^2}$ in (40).

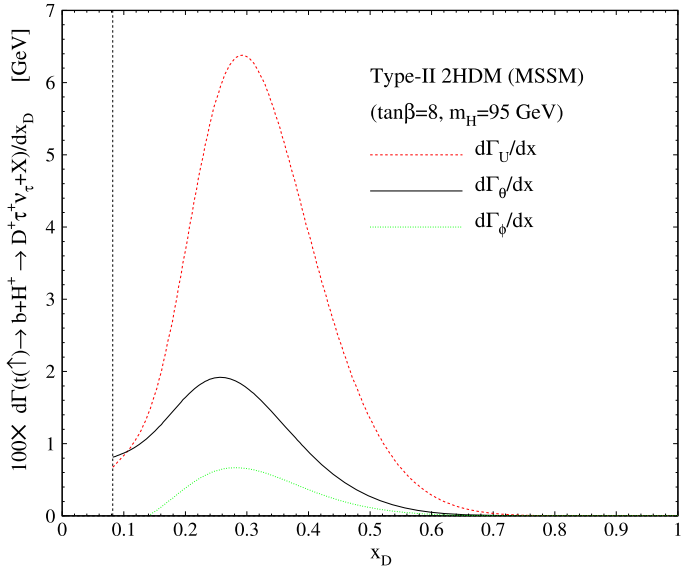

Fig. 7. $x_D$ spectrum for $D^+$-meson in top decay at NLO, with the hadronization modeled according to the Bowler model (54). The unpolarized (dashed line), the polar (solid line) and the azimuthal (dotted line) contributions are shown. Here, we set $\tan\beta = 8$ and $m_{H^+} = 95$ GeV.

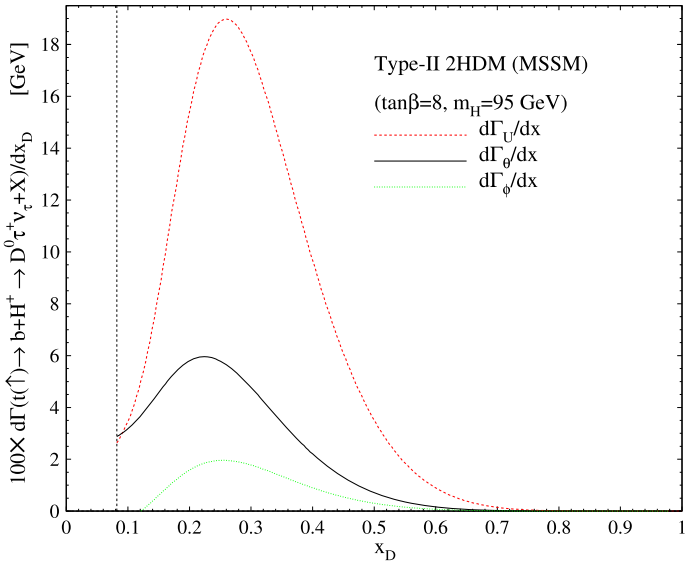

Fig. 8. As in Fig. 7 but for $D^0$-meson. Threshold occurs at $x_B = 0.08$.

Our formalism elaborated in this manuscript can be also extended to the production of hadron species other than bottom- and charmed-flavored hadrons, such as pions, kaons and protons, etc., using the nonperturbative $(b, g) \to \pi/K/P$ FFs extracted in [50], relying on their universality and scaling violations.

## 6. Conclusions

In a general 2HDM, the main production mode of light charged Higgs bosons is through the top quark decay, $t \to bH^+$ followed by $H^+ \to \tau^+ \nu_\tau$. On the other hand, bottom quarks hadronize before they decay, therefore, the study of energy distribution of hadrons inclusively produced in top decays is of prime importance at the LHC. This study is proposed as a new channel to indirect search for the light charged Higgs bosons.

In this work, by working in the NWA framework we studied the $\mathcal{O}(\alpha_s)$ spin-dependent energy spectrum of the charmed- and bottom-flavored hadrons produced through polarized top quark decays; $t(\uparrow) \to bH^+ \to B/D(\tau^+ \nu_\tau) + X$. The outgoing mesons are identified by a displaced decay vertex associated which charged lepton tracks.

In order to make our predictions for the energy spectrum of observed mesons we, for the first time, calculated the analytic expressions for the $\mathcal{O}(\alpha_s)$ corrections to the angular correlation functions $d\tilde{\Gamma}_\theta/dx_i$ and $d\tilde{\Gamma}_\phi/dx_i$ ($i = b, g$) in a specific helicity coordinate system where the event plane lies in the $(\hat{x}, \hat{z})$ plane and the b-quark three-momentum is considered along the $\hat{z}$-axis. In this frame, the polarization vector of top quark is evaluated relative to the positive $\hat{z}$-axis. These quantities are required to calculate the spin-dependent differential width $d\Gamma/(dx_B d\cos\theta_P d\phi_P)$ of the process $t(\uparrow) \to B/D(\tau^+ \nu_\tau) + X$. Since, highly polarized top quarks will become available at hadron colliders through single top production processes, which occur at the 33% level of the $t\bar{t}$ pair production rate [8], and in top quark pairs produced in future linear $e^+ e^-$-colliders [51] these studies can be considered as an indirect probe to search for the charged Higss bosons. In fact, any deviation of the B/D-meson energy spectrum from the SM predictions [16] can be considered as a signal for the existence of charged Higgs at the LHC.

Our analytical results can be also applied for the exotic decay channel $t(\uparrow) \to b(\to B + jet) + H^+(\to AW^+/HW^+)$, where $A$ and $H$ are the physical neutral Higgs bosons predicted in the 2HDM. This channel can reach a sizable branching fraction at low $\tan\beta$ [9]. Due to experimental challenges at low energies, such a light neutral Higgs has not been fully excluded yet.

As experimental considerations, two following points are discussed:

1)- In the decay modes $t \to bH^+/W^+$ followed by $H^+/W^+ \to \tau^+ \nu_\tau$, the tau leptons arising from the decays $W^+ \to \tau^+ \nu_\tau$ and $H^+ \to \tau^+ \nu_\tau$ are predominantly left- and right-polarized, respectively. The polarization of the $\tau^+$ influences the energy distributions of the decay products in the subsequent decays of the $\tau^+ \to \pi^+ \bar{\nu}_\tau, \rho^+ \bar{\nu}_\tau, l^+ \nu_l \bar{\nu}_\tau$ and, in conclusion, it influences the energy distribution of mesons produced through the b-quark hadronization. Therefore, the energy spectrum of (for example) B-mesons inclusively produced in the decay chain $t \to b(W^+, H^+) \to B(\tau^+ \nu_\tau) + X$ followed by $\tau^+ \to \pi^+ \bar{\nu}_\tau, \rho^+ \bar{\nu}_\tau, l^+ \nu_l \bar{\nu}_\tau$ can also help in searching for the induced charged-Higgs effects at the Tevatron and the LHC. Strategies to enhance the $H^\pm$-induced effects in the decay $t \to bW^+ \to b(\tau^+ \nu_\tau)$, based on the polarization of the $\tau^+$ have been discusses at length in Ref. [13] and references therein.

2)- Concerning the possible backgrounds on the proposed channels to search for charged Higgses ($t\bar{t} \to W^\pm H^\mp b\bar{b}$ and $t\bar{t} \to H^\pm H^\mp b\bar{b}$) it should be noted that the main background for light charged Higgs events is the SM top pair production process ($t\bar{t} \to W^\pm W^\mp b\bar{b}$). Since, both signal and background are produced from the same hard event, distinguishing beyond SM (BSM) top quark decay needs careful selection criteria; the most important of which originates from the fact that the charged Higgs decay products acquire predominantly higher four-momenta due to the higher charged Higgs mass compared to that of W-boson in the SM top decay $t \to bW^+$. Comparing this decay with the BSM one, $t \to bH^+$, and by assuming a final decay to the tau lepton for both W- and H-bosons, the tau lepton in SM background is produced with softer kinematic distribution than the one in signal events. The produced hadrons in top decays also have different kinematic distributions. This feature and the differences between the tau lepton kinematic distributions are clues for signal identification for further studies like the one presented in this paper. The overall selection strategy should be similar to what has been used at LHC charged Higgs observation and therefore follows the LHC analyses aiming at light charged Higgs observation at the currently open part of the $m_{H^\pm} - \tan\beta$ parameters space. By increasing the data set and $\tan\beta$ the signal receives more purity and resolution and in conclusion the error of measurements would be less.